# Implementation of an Improved Coulomb-Counting Algorithm Based on a Piecewise SOC-OCV Relationship for SOC Estimation of Li-Ion Battery

Ines Baccouche*, **, ‡, Sabeur Jemmali*, Asma Mlayah*, Bilal Manai***, Najoua Essoukri Ben Amara*

* LATIS-Laboratory of Advanced Technology and Intelligent Systems, ENISo, Sousse University, 4002 Sousse, Tunisia

** National School of Engineering of Monastir, Monastir University, Ibn El Jazzar 5019, Tunisia

*** IntelliBatteries/Yaslamen, Technopole of Sousse, BP24 Sousse Corniche 4059, Tunisia

(ines.baccouche@hotmail.fr, sabeur.jemmali@aist.enst.fr, assoum.mlayah@live.fr, bilalmanai@gmail.com, benamaranajwa@gmail.com)

‡ Corresponding Author; Ines Baccouche, ENISo BP 526, 4002 Sousse, Tel: +21621536422, ines.baccouche@hotmail.fr

**Abstract-** Considering the expanding use of embedded devices equipped with rechargeable batteries, especially Li-ion batteries that have higher power and energy density, the battery management system is becoming increasingly important. In fact, the estimation accuracy of the amount of the remaining charges is critical as it affects the device operational autonomy. Therefore, the battery State-Of-Charge (SOC) is defined to indicate its estimated available charge. In this paper, a solution is proposed for Li-ion battery SOC estimation based on an enhanced Coulomb-counting algorithm to be implemented for multimedia applications. However, the Coulomb-counting algorithm suffers from cumulative errors due to the initial SOC and the errors of measurements uncertainties, therefore to overcome these limitations, we use the Open-Circuit Voltage (OCV), thus having a piecewise linear SOC-OCV relationship and performing periodic re-calibration of the battery capacity. This solution is implemented and validated on a hardware platform based on the PIC18F MCU family. The measured results are correlated with the theoretical ones; they have shown a reliable estimation since accuracy is less than 2%.

## 1. Introduction

Currently, we are living in big developing in electronic applications, especially multimedia ones like smartphones, tablets and PCs, where portability remains the most important advantage. To strengthen the effectiveness of these mobile devices, a challenge of autonomy rises to make them reliable as long as possible. Hence, to ensure a permanent energy supply, we resort to rechargeable batteries. Actually, there are several kinds of rechargeable batteries used in industry: Lead-acid, Ni-MH, Ni-Cd, and Li-ion [1,2]. Nevertheless, the portable electronic devices tend to be more compact and lighter, so batteries based on the Li-ion technology seem to be more adequate than other batteries, thanks to their good characteristics; namely high energy density, high voltage, the important number of charge/discharge cycles, and safety, as reported in [2]. Besides, the state of the battery needs to be accurately controlled by algorithms designed to be embedded in the Battery Management System (BMS) in order to ensure a better use and a long battery life. Consequently, the concern of monitoring the remaining capacity is the most crucial task in BMS. Accuracy in State of Charge (SOC) estimation gives the precise energy available in the battery, easing the application control, avoiding irreversible damage to the battery's internal structure and ensuring optimal utilization. In literature, a great number of SOC estimation methods are available [3,4,5]: electrochemical [6], book-keeping [7,8,9,10], model-based [11,12] and black-box, also called data-oriented [13,14]. Each of these methods has strengths and weaknesses, and our goal is to develop a solution offering a compromise between accuracy and simplicity since we are targeting a hardware implementation on the BMS of a Li-ion battery.

Indeed, in this paper, we propose an efficient SOC estimation algorithm based on the Coulomb-counting algorithm which is a book-keeping approach. Using a piecewise linear relationship mapping between SOC and the Open Circuit Voltage (OCV) we try to overcome the drawbacks of this approach so as to improve its accuracy.

Actually, the implementation of the proposed algorithm on a hardware platform based on a microcontroller has led to a reliable system for SOC estimation of Li-ion batteries in term of precision and computation at the ambient temperature.

The remainder of this paper is structured as follows. Section 2 is dedicated to the related work. In section 3 we introduce the suggested algorithm based on Coulomb-counting and a novel SOC-OCV relationship and we detail each step of the method. In section 4, we explain the hardware implementation and the obtained results, and finally we present the conclusion and some perspectives for this work in section 5.

## 2. Related work

It has always been a big concern to estimate the SOC for energy storage devices. The estimation accuracy of SOC does not only give an information about the remaining useful capacity, but also indicates the charge and discharge strategies, which have a significant impact on the battery. Thus, a Li-ion battery may have different capacities due to aging, ambient temperature and self-discharge effects. Several methods for estimating SOC have been introduced in the literature. In this paper four main categories of SOC estimation methods are identified as presented in Table 1. The electrochemical methods are reported as high accurate because they deal with the internal properties of the battery such as lithium ions dimensions and the electrolyte concentration [15], but they are difficult to implement since it is not evident to access to the chemical structure in on-line monitoring systems. The model-based methods use various models and algorithms to calculate the SOC. Actually, they require an equivalent model used to simulate the battery behaviors. Thus many models types were suggested in literature [16]. These methods also require an adaptive algorithm generally based on state observers such as Kalman filter [17,18,19,20], Luenberger observer [17,21] and others [12,22,23]. Accordingly, their accuracy depends on the efficiency of the battery model and the precision of its characterized parameters. The data-oriented methods, essentially based on artificial intelligence algorithms like neural network [13,20] and fuzzy logic [14,20,24], estimate the SOC accurately for all kinds of batteries considered as black-boxes without the need for any information about the internal behaviors. These methods require a large number of training data. Therefore, they need powerful and costly processing and their effectiveness depends on the accuracy of the learning data. The book-keeping methods also known as Coulomb-counting, consist of a temporal integrating of the battery current during charge and discharge. The accuracy of these methods [7,8,9] is strongly dependent on the precison of current sensors.These non-model-based methods may accumulate errors caused by measurements, possible embedded noise and an inaccurate initial SOC.

Table 1. Review of SOC estimation methods

| SOC estimation method | Characteristics |
|---|---|
| **Electrochemical** <br> - Impedance spectroscopy [15] | - High accuracy <br> - Time consuming <br> - Hard to implement |
| **Model-Based** <br> - Kalman filter, extended Kalman filter [18,25,26] <br> - Lunenberger observer [21] <br> - Proportional integral observer [23] <br> - Sliding mode observer [12,22] | - Good precision <br> - Accuracy depends on the precision of the battery model . <br> - Not easy to implement |
| **Data-oriented** <br> - Neural Network [13] <br> - Fuzzy Logic [14] <br> - SVM [27] | - High accuracy <br> - Hard to implement <br> - Accuracy depends on the training data |
| **Book-keeping** <br> - Coulomb-counting [7,8] | - Average precision <br> - Simple to implement <br> - Cumulative errors <br> - Accuracy depends on sensors measurement |

Given that the aim of this work is to develop an embedded monitoring system, and considering that the data-oriented and the model-based methods are restrictive in terms of hardware implementation--since they require large data memory storage and heavy computation to describe the battery behaviors [5,28]-- we opt for the Coulombmetric counting method. Indeed, Coulomb-counting is only based on direct measurement, so it is not hard to implement and gives enough precision of the SOC estimation in multimedia applications.

Coulomb-counting based algorithms are often used as a core technology for battery SOC estimation in BMS. They express the SOC as the ratio of available capacity to the nominal one. The remaining capacity in a battery can be calculated by measuring the current flow rate (charging/discharging) and integrating it over the time interval $\Delta\tau$.

The common equation to calculate the SOC is given by Eq. (1), where the $SOC_0$ represents the initial SOC, $I_{bat}$ represents the current across the battery and, $Q_{rated}$ is the nominal capacity of the battery.

$$SOC = SOC_0 + \frac{\int_{t_0}^{t_0+\tau} I_{bat}\Delta\tau}{Q_{rated}} *100 \qquad (1)$$

However, this book-keeping method suffers from cumulative errors problem and the inaccuracy of the initial SOC estimation. Several researchers have reported these limitations and have proposed some solutions to solve them. For example, in [7], to avoid cumulative errors, a coefficient of Coulombic efficiency was added to the Eq. (1). In the same vein, the accuracy of the initial SOC affects severely the estimated SOC since $SOC_0$ is an additive term according to the definition in Eq. (1). Subsequently, we need to estimate $SOC_0$ with rigor and make the access to the initial SOC easier, for that the OCV-SOC model is generally used; because knowing the OCV value leads to having the instantaneous SOC owing to a well-known mapping between OCV and SOC. For this reason, the OCV-SOC model is widely utilized in a lot of works for battery characterization and monitoring [6,4,29,30]. The OCV-SOC mapping function can be implemented either as lookup tables [31,32] or an analytical expression [33,34,35].

The latter solution gives a better computational efficiency and can be generalized for all types of Li-ion batteries, unlike the lookup tables which are specific to each battery and are still too heavy to be implemented [33]. The analytical expression of the OCV-SOC curve has been differently defined in literature [25]. Figure 1 displays the typical curve of a Li-ion OCV-SOC. model.

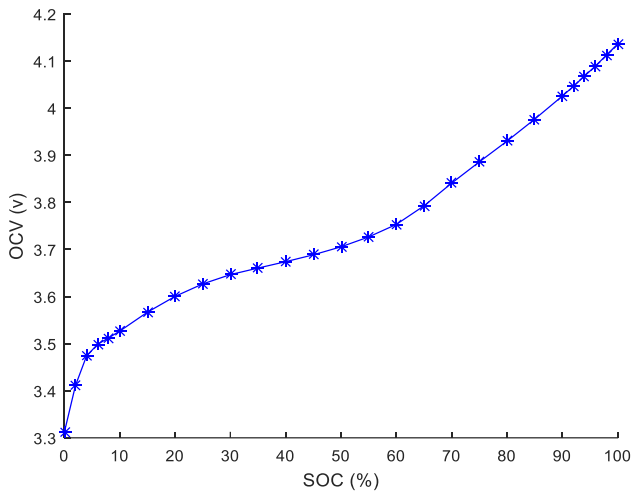

**Fig.1.** Typical OCV-SOC curve for Li-ion batteries.

Various works tried to give a fitting equation to this curve. In fact, in [5, 26] it was reported that the OCV-SOC relationship can be approximated to a linear segment which is not correlated with the reality but rather gives an acceptable accuracy and eases the implementation. Many other forms of OCV-SOC fitting were proposed which respect the nonlinear behavior of the battery and give a high accuracy. However, they have been considered to be unsuitable for hardware implementation [33]. In order to combine both accuracy and low complexity, the OCV-SOC curve was expressed as a piecewise linear system, as in [36], to have a reliable initial SOC estimation.

In this paper we propose an efficient Coulomb-counting algorithm of a SOC estimation for Li-ion batteries within a hardware platform based on microcontroller which can be used as a real-time tool for multimedia applications. The accuracy of this algorithm is guaranteed by using a piecewise SOC-OCV model to identify the intial SOC, a periodical recalibration to overcome cumulative errors due to an eventual inaccuracy of sensors and a temperature monitoring using an adaptive coefficient. The experiments prove that the suggested method is reliable and stable, and they result in obtaining SOC with less than 2% of error.

## 3. Proposed method for SOC estimation

Coulomb-counting is based on exploiting the Eq. (1) by quantifying the charge delivered by the battery through sensing its input and output current [7,37]. Yet, there are some inefficiencies when using this method. First of all, the access to the initial SOC is not guaranteed. Secondly, self-discharge may distort the real SOC value after a long storage period and finally the reference capacity $Q_{rated}$ must be updated in terms of battery ageing. Furthermore an SOC estimation is performed at varying ambient temperature which should be taken into account.

The literature shows enhanced Coulomb-counting algorithms which allow the determination of the initial SOC using the SOC-OCV curve, takes into account self-discharge losses and performs a recalibration at fully and empty stages.

Figure 2 gives the flowchart of the propopsed algorithm used to manage the battery monitoring by switching from a battery operating mode to another in accordance with transition conditions.

The estimation of the SOC of a Li-ion battery by the developed method is based on monitoring the voltage $V_{bat}$ the current $I_{bat}$ and the temperature T°. The operation mode of the battery is recognized by the direction of the current through the battery system. When the battery is in an open circuit mode, $I_{bat} = 0$, a compensation of self-discharge losses will be considered as will be explained in the sub-section 3.4. The information needed to perform the monitoring are the measurement of the battery voltage and the current flowing through it. Coulomb counter is used to track the SOC when battery is charging, discharging and self-discharging. The amount of charges in an operating period is obtained by a temporal integration of a measured charging/discharging current $I_{bat}$ as expressed in Eq. (2).

$$\Delta Q = \int_{t}^{t+\tau} I_{bat} dt \qquad (2)$$

This variation that will be used in Eq. (2) will be negative if the battery is in discharge and will be positive if in charge.

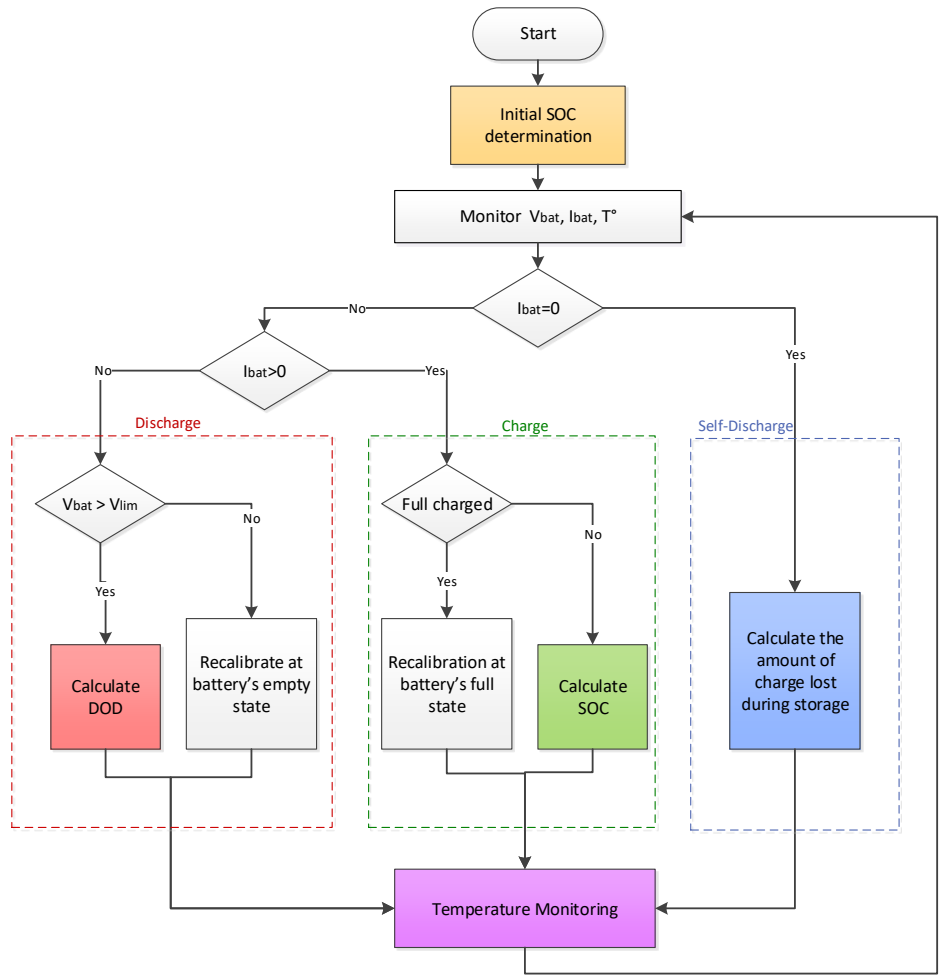

**Fig. 2.** Flowchart of the proposed algorithm.

*3.1. Initial SOC determination*

Several studies have been conducted to remedy to the problem of accurate estimation of the initial SOC [36]. The most common technique is to use a bijective OCV-SOC function that relates the open circuit voltage to its corresponding SOC value as detailed in the in section 2. This curve is actually determined experimentally by the OCV test by applying a pulse load on the Li-ion battery, then the battery reach an equilibrium where the voltage is extracted in every 5% of Depth Of Discharge (DOD) [38].

In this proposed algorithm, we start by determining the $SOC_0$ from measuring the initial OCV, for that we consider the inverse function noted SOC-OCV relationship of Li-ion battery which can be approximated to piecewise linear curve. As demonstrated in Fig.3, the curve is divided into eight segments [25], and each segment can be expressed as a linear relation as in Eq. (3):

$$SOC = f(OCV) = a*OCV - b \qquad (3)$$

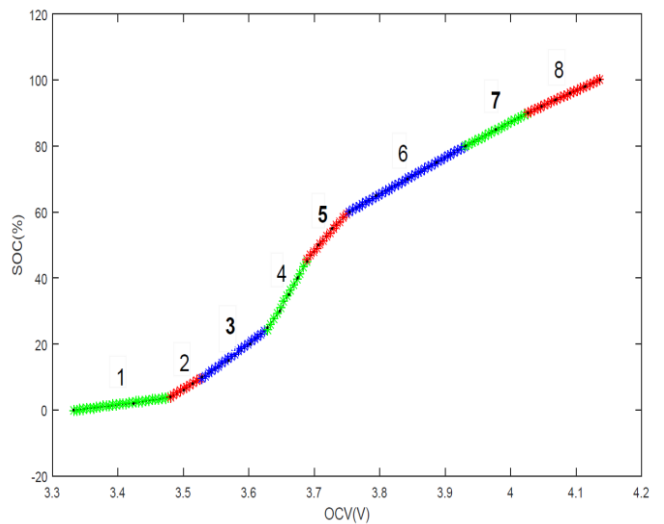

**Fig.3.** The proposed piecewise linear relationship of SOC-OCV.

For each segment we are varying coefficients a and b according to OCV intervals. The coefficients for each segment are given in Table 2, with their voltage range.

Table 2. Piecewise linear relationship of SOC-OCV at 25°C

| Segment | 1 | 2 | 3 | 4 | 5 | 6 | 7 | 8 |
|---|---|---|---|---|---|---|---|---|
| Voltage Range (V) | [3.3; 3.452] | [3.452; 3.508] | [3.508; 3.595] | [3.595; 3.676] | [3.676; 3.739] | [3.739; 3.967] | [3.967; 4.039] | [4.039; 4.132] |
| A | 26.55 | 125 | 149 | 344 | 229.5 | 111.9 | 104.8 | 90.61 |
| B | 88.6 | 431.1 | 516.1 | 1225 | 800.9 | 359.9 | 332 | 274.7 |

### 3.2. Charge Mode

The typical charge procedure of a Li-ion battery consists of constant-current constant voltage (CC-CV) process [29]. The Coulomb counter is presented by $Q_{gained}$ as expressed in Eq.(4), which represents the quantity of charge accumulated during an operating period equal to τ.

$$Q_{gained}(t+\tau) = Q_{gained}(t) + \Delta Q \quad (4)$$

Thus, the variation of the state of charge gained in this same operating period is obtained by the Eq. (5) [25].

$$\Delta SOC(t+\tau) = \frac{Q_{gained}(t+\tau)}{Q_{rated}} *100 \quad (5)$$

By accumulating the previous state of charge indication and the obtained one, we can have the instantaneous value of SOC as expressed in (6):

$$SOC(t+\tau) = SOC(t) + \Delta SOC(t+\tau) \quad (6)$$

Knowing the relationship indicated in Eq. (7) the value of DOD is updated in every charging operation in order to get it back in every switching to discharging mode.

$$SOC(t) + DOD(t) = 100\% \quad (7)$$

### 3.3. Discharge mode

Figure 4 shows the typical voltage curves when a Li-ion battery is discharged by different C-rate.

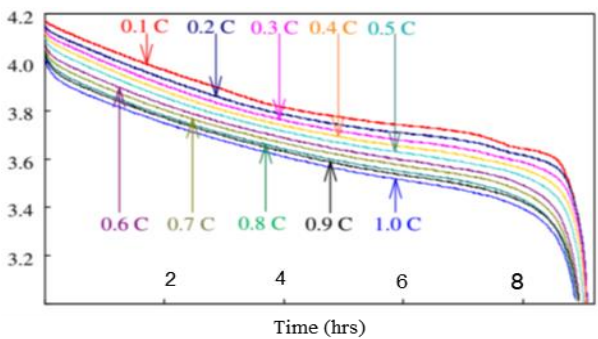

**Fig.4.** Typical discharge profiles of Li-ion battery under different currents (C-rate) [7]

When the current $I_{bat}$ is negative, then the battery is in the discharging mode. In this mode, the Coulomb counter is presented by $Q_{lost}$, which represents the amount of charges losses in the operating period τ as expressed in Eq. (8).

$$Q_{lost}(t+\tau) = Q_{lost}(t) + \Delta Q \quad (8)$$

The value of DOD is calculated according to Eq. (9)

$$DOD(t+\tau) = DOD(t) + \Delta DOD(t+\tau) \quad (9)$$

Where : $\Delta DOD(t+\tau) = \dfrac{Q_{lost}(t+\tau)}{Q_{rated}} *100$

When the initial value $SOC_0$ is known as explained in section 3.1, the instantaneous SOC is evaluated by integrating battery current over time, as depicted in Eq. (1). Hence, SOC indication is done by applying the Eq. (7).

### 3.4. Self-Discharge Mode

At the battery storage periods, considering that Li-ion batteries reach 5% rate of self-discharge per month [2], the amount of charge losses per hour is calculated; this amount is designed by the constant, $q_{per/hour}$. Then, the quantity of charge dissipated in this phase $Q_{oc}$ is calculated by Eq. (10), representing the cumulative losses during the storage hours.

$$Q_{oc}(h+1) = Q_{oc}(h) + q_{per/hour} \quad (10)$$

where $h$ is the hour of storage

This value will be added to the amount of charges lost $Q_{lost}$ during discharge mode and subtracted from the amount of charge accumulated $Q_{gained}$ in the charge mode as expressed in Eq. (11):

$$\begin{cases} Q_{lost} = Q_{lost} + Q_{oc} \\ Q_{tot} = Q_{tot} - Q_{oc} \end{cases} \quad (11)$$

This compensation of self-discharging loss is made at each open circuit period and before switching to another operating battery's mode.

### 3.5. Temperature Monitoring

Different studies have highlighted that the temperature has an important effect on the battery behavior [29]. In fact, at low values the released capacity from the battery decreases dramatically as illustrated in Fig.5. One can see that the capacity is about 80% when the temperature is about 5°C and reaches 60% when T°= -10°C.

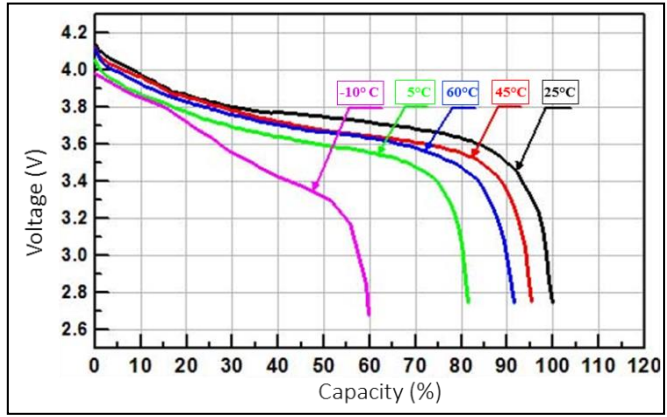

**Fig.5.** Typical discharge profiles of Li-ion battery under various temperatures [39,40].

The effect of temperature is also detected at the level of the SOC-OCV model especially in the SOC range between 0-30% as shown in Fig.6.

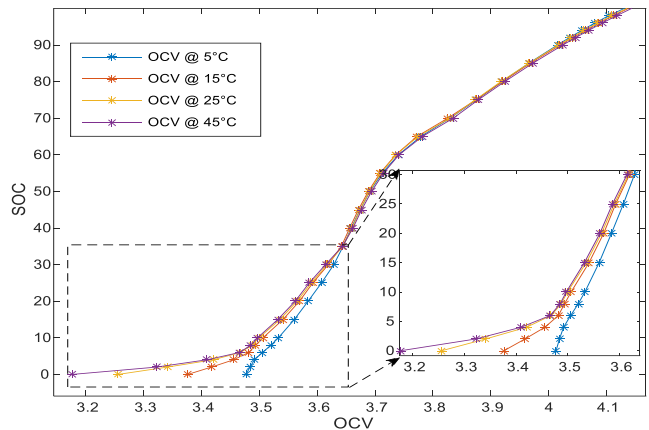

**Fig.6.** Variation in SOC-OCV model at various temperatures

According to Fig.6 the OCV model is the same for all temperature values except for those under 15°C. At the initial SOC determination phase, we are considering the effect of the temperature variation. Indeed, when the temperature value is in the range [5°- 15°C], we only take into account the last seven segments of the OCV battery model presented in Table 2 (segments 2 to 8). Furthermore, for temperatures under 5°C we consider only the last six segments, thus ignoring the first two segments.

With respect to the Fig.5, in order to consider the ambient temperature variation, we propose to use a variant coefficient α to weight the SOC value such to adapt it to the correspondent external temperature (c.f. Table 3.).

Table 3. The coefficient α at various temperatures

| Temperature | 60°C | 45°C | 25°C | 5°C | -10°C |
|---|---|---|---|---|---|
| α | 0.9 | 0.95 | 1 | 0.8 | 0.6 |

For intermediate temperature values, the coefficient α is determined by considering a linear interpolation as shown in Fig.7.

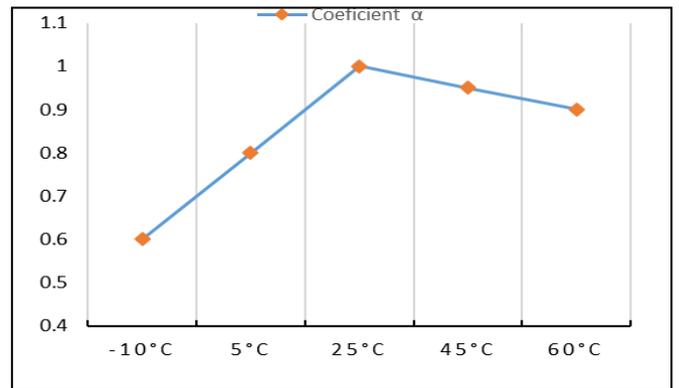

**Fig.7**. The linear interpolation of the temperature coefficient

For a hardware implementation concern we reduced the variation of the coefficient α to five main intervals as given in (12).

$$\alpha = \begin{cases} 0.5 & T° < -10 \\ 0.6 & -10 \leq T° < 5 \\ 0.8 & 5 \leq T° < 25 \\ 1 & 25 \leq T° < 45 \\ 0.9 & 45 \leq T° < 60 \end{cases} \quad (12)$$

### 4. Hardware Implementation

The adopted Coulomb-counting algorithm for SOC estimation was implemented on a hardware platform. The monitoring functions include the measurement of battery voltage and the current flowing through it by the 10 bits Analog to Digital Converter MCU PIC18F.

### 4.1. Measurement

The battery parameters detection is one of the most important issue in control and management of a BMS [6]. It includes cell voltage measurement and current detection. Estimation of SOC and other battery states imposes high requirements on cell voltage precision [6]. The diagram presented in Fig.8 shows the different modules connection to the data acquisition system.

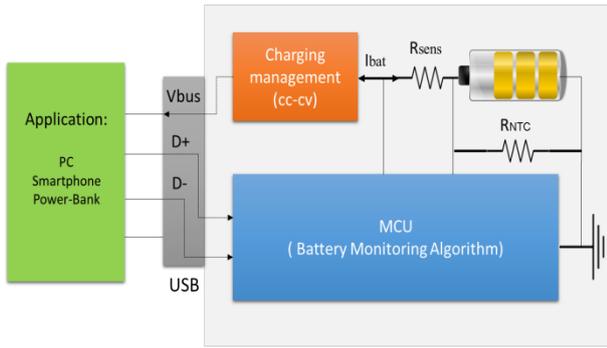

**Fig.8**. The basic circuit developed

In order to measure the voltage and the current with which the battery is charged or discharged, a sensing resistor $R_{sens}$ is placed in series with the battery.

The battery voltage $V_{bat}$ is obtained by measuring the voltage across it. The difference between the voltages at the two terminals of the resistor will derive the current value $I_{bat}$ through the sensing resistance. This resistance, which allows the current sensing in the battery gauge, should not be too high in order to avoid power dissipation during the charging at a constant current of 1C-rate, but it should still be enough so that the analog/digital converter in the micro controller can detect a voltage variation across it, which corresponds eventually to a 10 mA current, in order to detect the end of charging. In our case, a 10-bit converter is used to detect a voltage variation of 1mV. A 0.1 Ohms resistance can detect a current of 10 mA (for U = 1mV) and does not dissipate too much power while charging the maximum current value (P=100mW). The transmission of the battery electrical parameters is ensured by a USB connection between the multimedia application and the MCU PIC18F.

The ambient temperature is measured using a negative temperature coefficient resistor $R_{NTC}$. This thermistor is biased by IO of the MCU and connected close to the batterie as depicted in Fig 8.

To show the measurements, a graphical interface is used to display the electrical parameters of the battery: battery voltage: $V_{bat}$, voltage across $R_{sens}$ : $V_{pack+}$, battery current : $I_{bat}$, and charging voltage : $V_{bus}$ (Fig.9).

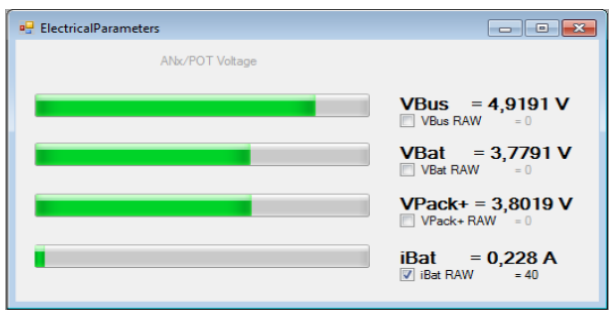

**Fig. 9.** Graphic interface of the measurement

*4.2. Application*

To validate the developed algorithm, some tests were achieved using a test card developed in the company IntelliBatteries. It consists of five main parts which are shown in Fig.10:

➢ The charging management ensuring cv-cc procedure.
➢ The MCU that contains the programmable card PIC18F.
➢ The NTC resistor that indicates the external ambient temperature
➢ The protection module that protects the battery against overload, voltage drops and power surges.
➢ The USB connection that ensures both the charging process and the data transmission from the multimedia application to the MCU.

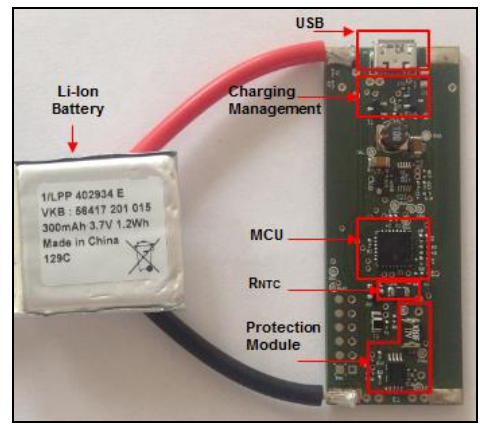

**Fig.10.** The test card

The monitoring algorithm is then implemented with embedded C language on the MCU. The diagram presented in Fig.11 includes a call-graph generated by the MPLAB software. It lists the functions used in monitoring the SOC of the battery.

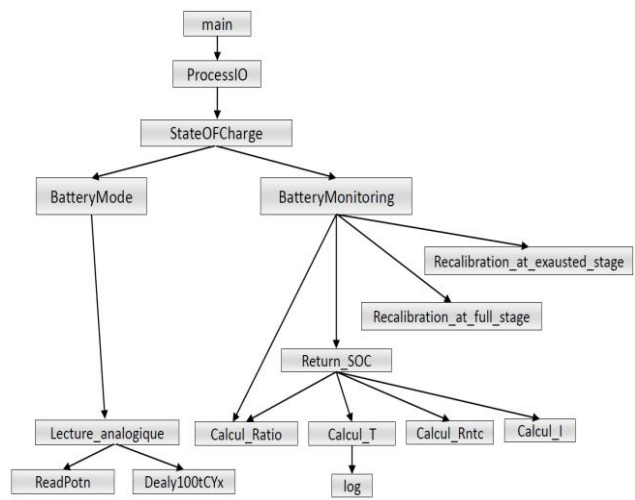

**Fig.11.** The callgraph of implementation of the proposed Coulomb-counting

*4.3. Results and Experimentations*

Once the integration of the monitoring algorithm is completed, we conduct some tests to validate the effectiveness of the developed system. A comparison between the Coulomb-counting algorithm simulated by Matlab and the algorithm embedded on the MCU is performed to prove the reliability of our proposed solution. This test is realized during the charging mode of two Li-ion batteries connected in parallel with a total nominal capacity of 4400 mAh.

In Fig.12, the red curve represents the cumulated capacity obtained by Matlab simulation, whereas the blue one is the result of the same algorithm embedded on hardware platform. The difference detected between the two curves for the same algorithm with the same data is due to the fact that Matlab is a high-precision calculator, while the MCU has an ADC of a resolution of only 10 bits. Fig.12 proves the reliability of our developed system, since the two curves are nearly superposed and the error is almost 2% and does not exceed 5% as illustrated in Fig.13.

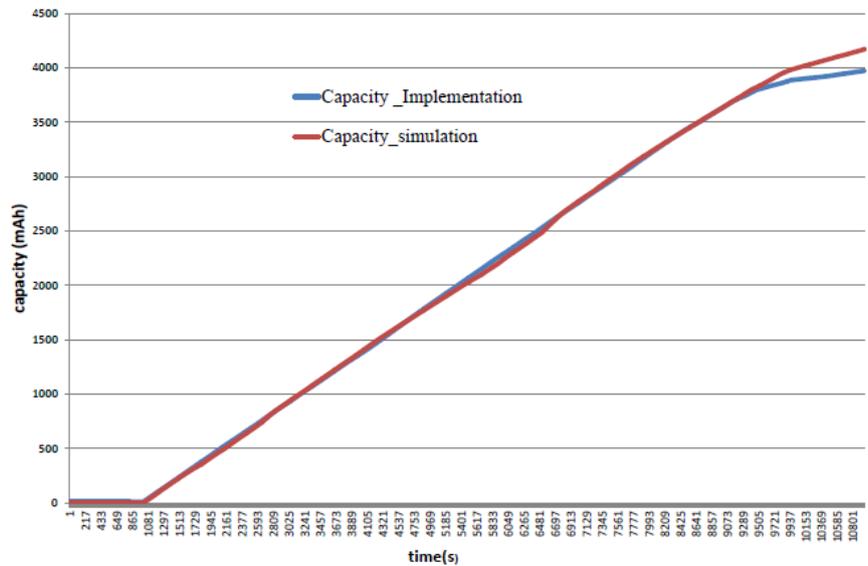

**Fig.12.** The instantaneous capacities obtained by both simulation and implementation

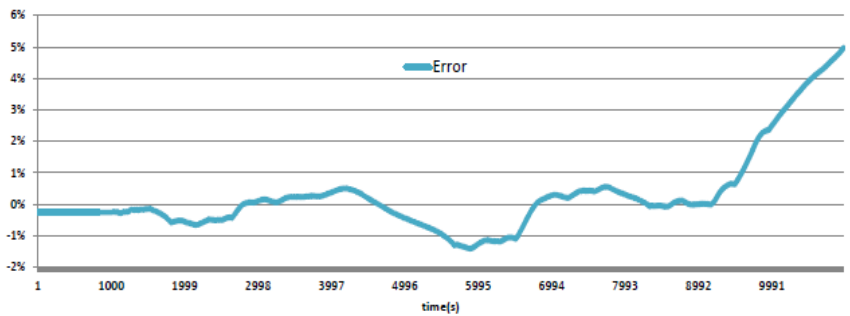

**Fig.13.** The error curve between simulation and implementation

## 5. Conclusion

In this paper, an improved Coulomb-counting method for the SOC estimation for Li-ion batteries has been presented. This algorithm has allowed the system to operate in safe conditions by respecting the secure operating area of the Li-ion batteries in charge and discharge modes. In order to enhance the SOC estimation accuracy, we have proposed a SOC-OCV model as a piecewise linear relationship composed of eight segments. This contribution has allowed overcoming the disadvantage of the uncertainty of the initial SOC value. Also a recalibration at both full and empty states of the battery has been carried out, based on updating the amount of the releasable capacity. The recalibration has contributed to decrease cumulative errors caused by the inaccuracy of sensors. These improvements added to the Coulomb-counting algorithm have enhanced the SOC estimation accuracy. Moreover, we have put forward an adaptation procedure of the calculated value of SOC using a weighting coefficient determined respecting various temperature ranges. A hardware implementation of this algorithm was performed on an MCU PIC18F and validated by several tests of charge and discharge. The proposed algorithm has been simple to implement since that the SOC-OCV has been a linear relationship and the coefficients are not too consuming in terms of memory storage. The hardware implementation has proven a good reliability and the error estimation of the instantaneous capacities has been only about 2%. This proposed system has been supported by IntelliBatteries company and it has been integrated on its products. Presently, we are developing and characterizing a new battery model that simulates the nonlinear behaviors of the Li-ion batteries, taking into account other constraints like the temperature effect, C-rate, and aging so as to enhance the accuracy and reliability of the SOC estimation.

**Acknowledgement**

This work is part of a PhD thesis supported by PASRI−MOBIDOC project, IntelliBatteries company and LATIS laboratory.